\documentclass[12pt]{JHEP3}
\usepackage{pb-diagram,lamsarrow,pb-lams,amsfonts,amssymb,amsthm}

\usepackage{amsmath,epsfig}
\def\be{\begin{equation}}
\def\ee{\end{equation}}
\def\ba{\begin{eqnarray}}
\def\ea{\end{eqnarray}}

\def\be{\begin{equation}}
\def\ee{\end{equation}}
\def\bea{\begin{eqnarray}}
\def\eea{\end{eqnarray}}

\def\yzero{\smash{\hbox{$y\kern-4pt\raise1pt\hbox{${}^\circ$}$}}}

\def\m{\mu}

\def\beq{\begin{equation}}
\def\eeq{\end{equation}}
\def\beqa{\begin{eqnarray}}
\def\eeqa{\end{eqnarray}}

\def\-{\hphantom{-}}

\def\s2{\frac{1}{\sqrt2}}

\def\beq{\begin{equation}}
\def\eeq{\end{equation}}
\def\beqa{\begin{eqnarray}}
\def\eeqa{\end{eqnarray}}

\def\IF{\relax{\rm I\kern-.18em F}}
\def\II{\relax{\rm I\kern-.18em I}}
\def\IP{\relax{\rm I\kern-.18em P}}
\def\IC{\relax\hbox{\kern.25em$\inbar\kern-.3em{\rm C}$}}
\def\IR{\relax{\rm I\kern-.18em R}}

\def\Dsl{\,\raise.15ex\hbox{/}\mkern-13.5mu D} 
\def\IZ{Z\kern-.4em  Z}

\title{Dualities as symmetries of the Supermembrane Theory}

\author{M. P. Garc\'\i a del Moral,\footnote{E-mail:
\emph{garciamormaria@uniovi.es}}\\
Departamento de F\'\i sica, Universidad de Oviedo, Avda Calvo
Sotelo S/n. Oviedo, Espa\~na}
\abstract{ In this note I review the role played by dualities in the Supermembrane Theory compactified on a torus. Supermembrane theory realize S, T, so U-duality, as exact symmetries of the theory. There are two well defined sectors:  with and without central charges. Both sectors have the $SL(2,Z)_{T^2}\times Z_2$ U-duality group in 9D as a symmetry of the theory, but the supermembrane with central charges, also exhibits an extra $SL(2,Z)_{\Sigma}$ symmetry of the theory associated to the diffeomorphisms not connected with the identity changing the homology basis. The $Z_2$ symmetry is T-duality transformation that acts on the supermembrane and generalize that of the string. This T-duality transformation acts locally and also globally on the supermembrane theory. It has a natural description in terms of the cohomology of the base manifold and the homology of the target torus. When only string-like states are preserved, one recovers both type II string theories in 9D. If one also consider properly limit of the T-duality transformation characterizing the supermembrane, the usual closed string T-duality transformation between the type IIA and type IIB mass operators is recovered.}

\preprint{FPAUO-18/11}

\begin{document}
\section{Introduction}
In this note we review recent, and not so recent results that together with my collaborators, we have found in different papers \cite{sl2z},\cite{gmpr2},\cite{torus}. The purpose of this note is to give a unified picture of the dualities of the 2-torus compactified supermembrane. There are two well-defined sectors in the theory that have to be considered separately since they exhibit important physical differences. We have found that, however, both sectors realize S-duality, and T-duality , so U-duality, as exact symmetries of the theory although in a different way.

T-duality transformation at the worldsheet level were studied in \cite{cvetic-stelle}.
The relation of duality and M-theory was also analyzed in \cite{abou}. A topological analysis of T-duality in terms of toroidal bundles with monodromy form a mathematical perspective has been done in \cite{baraglia}. In \cite{4hull04} it was argued that a fundamental
formulation of string/M-theory should exist in which the T- and
U-duality symmetries are manifest from the start.  In particular, it
was argued that many massive, gauged supergravities cannot be
naturally embedded in string theory without such a framework
\cite{stw}, \cite{6hullreid},\cite{reids},
 \cite{7samtleben}.

Double field theory has become a interesting arena to
try to realize in a bottom-up approach, some of  the properties of
string theory. It is a global approach that describe sigma models
with double coordinates on a $T^{2d}$ torus fibrations such that the
transition functions will be evaluated in the T-duality group
$O(d,d,\mathbb{Z})$.There is evidence that string theory can be consistently defined in
non-geometric backgrounds in which the transition functions between
coordinate patches involve not only diffeomorphisms and gauge
transformations but also duality transformations
\cite{4hull04}.
 To this end it is relevant to understand the global description of dualities in terms of bundles. Some global aspects of T-duality in String theory formerly analyzed in \cite{lozano}, were more recently realized in this context  by
\cite{24hull-tfolds}. The type II realization has been done recently
in \cite{osb,os}. The proposed actions are such that they are invariant
under duality transformations.

In this
type of compactifications, T-folds are constructed by using strings
formulated on a doubled torus $T^{2n}$ with n-coordinates conjugate
to the momenta and the other n-coordinates conjugate to the winding
modes \cite{4hull04}, plus a constraint to guarantee the correct
number of propagating degrees of freedom. 
  Such backgrounds can arise from
compactifications with duality twists \cite{dh2} or from
acting on geometric backgrounds with fluxes with T-duality
 \cite{4hull04}, \cite{stw}.  In special cases, the
compactifications with duality twists are equivalent to asymmetric
orbifolds which can give consistent string backgrounds
\cite{12flournoyw}, \cite{13kawai},\cite{18cederwall}, \cite{dhgt}. Examples of
generalized T-folds can be obtained by constructing torus fibrations
over base manifolds with non-contractible cycles.
 The same type of approach can be also implemented in M-theory. There has been works done that realize effective duality invariant approach to M-theory by a U-duality group valued Scherk-Schwarz twist \cite{18cederwall,bmt} and are related with generalized geometry \cite{bgpw,dh2}. 
\newline

Our approach however will be different: we perform a top-down approach departing from a sector of M-theory: the supermembrane compactified on a torus. In the case of interest (nine noncompact dimensions) the conjectured group of U-dualities for type II string theories is the product: $SL(2,Z)\times Z_2$ \cite{hull-townsend}. Previous works on the role of dualities on the supermembrane were done by \cite{aldabe1},\cite{aldabe2},\cite{russo}.\newline
The compactified supermembrane on a target space $M_9\times T^2$ may be formulated in terms of sections on symplectic torus bundles \cite{gmmpr}. There are two well-defined sectors: one in which a topological condition due to an irreducible wrapping is imposed, that corresponds to the so-called supermembrane with central charges \cite{mor}, and a second one on which the previous condition vanishes. While the first one can be globally formulated in terms of sections on symplectic torus bundles  with $SL(2,\mathbb{Z})$ monodromy \cite{gmmpr} and at low energies corresponds to the type II gauged supergravities in 9D \cite{gmpr2} through a gauging mechanism that acts on the structure of the bundles \cite{sculpting}, the second one corresponds to the formulation on trivial symplectic torus  bundles \cite{gmpr2} associated to the maximal supergravity in 9D \cite{julia}. Physically the two sectors have very different properties among which, the most relevant one, is that the regularized supermembrane with central charges has discrete spectrum \cite{bgmr}-\cite{bgmr3} -so it is a well-defined quantum object- in distinction with the compactified supermembrane with vanishing central charge condition \cite{dwpp}. Due to this very relevant property it has become sensible to study the supermembrane as a quantum object in lower dimensions \cite{joselen},\cite{bellorin}, in G2 compactifications \cite{g2} or for example its non-abelian extension \cite{gmr2}.

In this note we show that the supermembrane compactified on a torus is invariant under a S-duality and T-duality transformation of the supermembrane \cite{gmpr2},\cite{sl2z},\cite{torus}. More concretely  $SL(2,\mathbb{Z})$ as a symmetry associated to the 2-torus target space appears in both sector of the theory. However there is an extra $SL(2,Z)$ that only acts on the supermembrane witn central charges sector \cite{sl2z}.

In \cite{gmpr2} we  showed the existence of a new $Z_2$ symmetry
that plays the role of T-duality in the supermembrane interchanging the
winding and KK charges but leaving the Hamiltonian invariant. T-duality becomes an exact symmetry of the
symplectic torus bundle description of the supermembrane by fixing its energy tension.
We will review show how in the String Theory limit, the T-duality transformation for the supermembrane becomes the standard T-duality transformation of the closed superstrings compactified on a circle \cite{torus}.\newline

The outline of this note is the following: In section 2, we review the compactified supermembrane on a torus and we establish the two well differenciated topological sectors, with and without central charges, or equivalently with non-trivial or trivial second cohomology class. In section 3, we show the two types of $SL(2,Z)$ symmetries that the compactified supermembrane can have. However only the supermembrane with central charges exhibits both. The compactified supermembrane with vnaishing second cohomology class only has the $SL(2,Z)$ symmetry associated to the 2-torus target space. In section 3, the T-duality transformation is introduced from a local and global approach. The differences in both sectors are pointed out.
It is shown that it realizes as a symmetry of the theory when a relation between the scale of energy and the moduli is established. In section 4, the string theory limits are discussed. When only string-like configurations of the supermembrane mass operator are preserved the perturbative type IIA and type IIB $SL(2,Z)$  mass spectrum are recovered. If in distinction, a T-dual transformation is performed on the M2 mass operator and then the string-like configurations are the only ones preserved then the nonperturbative $SL(2,Z)$ mas spectrum is obtained. If also a circle limit is imposed on the T-duality transformation then the usual T-duality for type II closed string is recovered. In section 5, conclusions and comments are presented.
\section{The Compactified Supermembrane in $M_9\times T^2$}
We consider now the compactified supermembrane embedded on a target space $M_9\times T^2$  where
$T^2$ is a flat torus. We consider
maps $X^m,X^r$  from $M_9\times T^2$ to the target space , where $X^m$ with $m=3,\dots,9$ are single valued maps onto the Minkowski sector of the target
space while $X^r$, with $r=1,2$ maps onto the $T^2$ compact sector of the
target. The winding condition corresponds to
\bea\begin{aligned}
\oint_{\mathcal{C}_{s}}dX^1=2\pi R (l_{s}+m_{s}Re\tau);\quad
\oint_{\mathcal{C}_{s}}dX^2=2\pi R m_{s}Im\tau;\quad
\oint_{\mathcal{C}_{s}}dX^m=0 \end{aligned}\eea where  $R,\tau$ are respectively the radius and the Teichmuller parameters of the 2-torus target-space, and $l_{s},m_{s}, s=1,2$, are integers.

The physical hamiltonian in the LCG is given by \begin{equation}\label{5}
\begin{aligned}
&\mathcal{H}= \int_{\Sigma}T_{11}^{-2/3}\sqrt{W}\left[\frac{1}{2}(\frac{P_{m}}{\sqrt{W}})^{2}+
\frac{1}{2}(\frac{P_{r}}{\sqrt{W}})^{2}
+\frac{T_{11}^{2}}{2}\{X^{r},X^{m}\}^{2}
+\frac{T_{11}^{2}}{4}\{X^{r},X^{s}\}^{2}\right]\\ \nonumber &+\int_{\Sigma}T_{11}^{-2/3}\sqrt{W}\left[\frac{T_{11}^{2}}{4}\{X^{m},X^{n}\}^{2}- \overline{\Psi}\Gamma_{-} \Gamma_{m} \{X^m,\Psi\}- \overline{\Psi}\Gamma_{-} \Gamma_{r} \{X^r,\Psi\}\right] \end{aligned}\end{equation} subject to the constraints 
\begin{equation}  \begin{aligned}\label{e2}
\phi_{1}:=& d(\frac{P_{m}}{\sqrt{W}}dX^{m}+\frac{P_{r}}{\sqrt{W}}dX^{r} -\overline{\Psi}\Gamma_{-}d\Psi)=0, \\
\phi_{2}:=& \oint_{C_{s}}(\frac{P_M}{\sqrt{W}}dX^M+\frac{P_{r}}{\sqrt{W}}dX^{r} -\overline{\Psi}\Gamma_{-}d\Psi)= 0,
   \end{aligned}
\end{equation}
associated with a residual symmetry of the theory: the infinite group of diffeomorphims preserving the Riemann basis $\Sigma$. So far, we have described the compactified supermembrane with no distinction between sectors. 
We now impose an extra topological
restriction on the winding maps \cite{mrt}: the irreducible winding constraint,
 \bea\label{central}
 \int_{\Sigma}dX^r\wedge dX^s=n\epsilon^{rs}Area(T^2)\quad n\ne 0, r,s=1,2.
 \eea
where $Area(T^2)=(2\pi R)^2 Im\tau$. \newline
There are two well-defined topological sectors classified according to this condition:
\begin{itemize}
\item{}
When this condition holds, $(n\ne 0)$ we refer to it as the supermembrane with central charge theory \cite{mor} and it  implies that the winding matrix $\mathbb{W}= \begin{pmatrix} l_{1}& l_{2}\\m_{1} & m_{2}\end{pmatrix}$ has  $det \mathbb{W}=n\ne 0$. Globally it corresponds to a sector, with nontrivial symplectic torus bundle with monodromy $\rho$ in $SL(2,Z)$, characterized by having $H^2(\Sigma,\mathbb{Z_{\rho}})\ne 0$ \cite{gmmpr},\cite{gmpr2}. \newline
\item{}
When this condition vanishes ($n=0$), this corresponds to a sector that we will call from now on compactified M2, $n=0$. Globally it corresponds to a trivial symplectic torus bundle over the base,  lets choose for symplicity the flat 2-torus $\Sigma_1$. It is characterized by having a trivial class of  $H^2(\Sigma_1,\mathbb{Z})$.
\end{itemize}

The Mass operator of the compactified supermembrane with winding and KK contribution \cite{sl2z}, \cite{schwarz}, is
\bea\label{28} Mass^{2}=T_{11}^{2}((2\pi R)^{2}n Im \tau)^{2}+
\frac{1}{R^{2}}((m_{1}^{2}+(\frac{m\vert q\tau-p\vert}{R Im\tau})+ T_{11}^{2/3}H\eea where the $H$ is defined in terms of the above hamiltonian $\mathcal{H}$ once the winding contribution has been extracted $H= \mathcal{H}-
T_{11}^{-2/3}\int_{\Sigma}\sqrt{W}\frac{T_{11}^{2}}{4}\{X^{r}_{h},X^{s}_{h}\}^{2}$ \cite{sl2z}. In the case of the sector of the  compactified M2, sector $n=0$, the winding contribution vanishes.

\section{The $SL(2,\mathbb{Z})$ Symmetry of the supermembrane }

The supermembrane is invariant under are preserving diffeomorphisms on the base manifold
 homotopic to the identity. This symmetry is realized by the first class constraints on the theory. Besides this standard symmetry of the supermembrane, the theory because of being compactified on a torus target is going to have a discrete symmetry $SL(2,Z)$.

In order to analyze the $SL(2,Z)$ symmetry of the supermembrane in detail \cite{sl2z}, we first perform a Hodge decomposition of the closed one-forms. We may decompose the closed one-forms $dX^{r}$ into
 \bea
dX^{r}=M_{s}^{r}d\widehat{X}^{s}+dA^{r}\quad r=1,2 \eea where
$d\widehat{X}^{s}, s=1,2$ is the basis of harmonic one-forms we have
already introduced, $dA^{r}$ are exact one-forms and $M_{s}^{r}$ are
constant coefficients.  This condition is satisfied provided
\bea M_{s}^{1}+iM_{s}^{2}=2\pi R (l_{s}+m_{s}\tau) \eea
Consequently, the most general expression for the maps $X^{r}$,
The general expression for the $dX$ maps is then
 \bea
dX=dX_{h}+dA
 \eea
The harmonic part of $dX$,
\bea\label{nt11}
dX_{h}=2\pi R [(m_1\tau+l_1) d\widehat{X}^{1}+(m_2\tau +l_2) d\widehat{X}^{2}].
\eea
with $l_{s}, m_{s}$, $s=1,2$, in principle of arbitrary
integers.

The two different sectors of the theory  although formally have the same expansion they differ and consequence of it, is the fact that they exhibit different properties of the harmonic sector depending whether $\mathbb{W}$ is equal or not to zero: 
\begin{itemize}
\item{}
For the  trivial symplectic torus bundle sector associated to the compactified M2, $n=0$, the harmonic sector contributions $X_r$ are not independent. This fact classically is responsible for the existence of the string-like spikes with zero energy that render the theory unstable, and together with supersymmetry at quantum level they are also responsible for the continuity of the spectrum \cite{dwln}. for this reason, the supermembrane with winding has continuous spectrum as said in \cite{dwpp}.
The Hodge decomposition can still be made but no gauge connection can be consistently defined in the theory since the coefficients behind the harmonic forms do not form a group as its determinant can be equal to zero. Although formally one can extract an exact 1-form this is not a gauge connection of the fibre since it is not possible to define properly the covariant derivative \cite{mrsymp},\cite{mor},\cite{gmr}.
\item{}
For the nontrivial symplectic torus bundle associated to the compactified M2 $n\ne 0$ the harmonic sector $X_r$ are independent. Classically There are not string-like spikes \cite{gmr} and the spectrum consists in purely isolated eigenvalues of finite multiplicity \cite{bgmr}-\cite{bgmr3} and then allows to define a covariant derivative, and $A$ is a connection over this bundle with structure group the symplectomorphisms group. $A$ is the gauge connection defined on the worldvolume associated to the monopole contribution \cite{mr}. 
   
There is a compatible election for $W$ on the
geometrical picture we have defined.  We
define \bea \label{8}
\sqrt{W}=\frac{1}{2}\epsilon_{rs}\partial_{a}\widehat{X}^{r}\partial_{b}\widehat{X}^{s}\epsilon^{ab},
\eea it is a regular density globally defined over $\Sigma$. It is
invariant under a change of the canonical basis of homology.
$X_h$ is a minimal immersion from $\Sigma$ to $T^{2}$ on the target,
moreover it is directly related to a holomorphic immersion of
$\Sigma$ onto $T^{2}$. The extension of the theory of supermembranes
restricted by the topological constraint to more general compact
sectors in the target space is directly related to the existence of
those holomorphic immersions \cite{bellorin}.
\end{itemize}
\subsection{The U-duality invariance}
The compactified supermembrane theory on a torus, -both sectors (with and without central charge)- is also invariant under the following transformation on the target torus $T^{2}$:
\bea \label{17}\tau &\to& \frac{a\tau+b}{c\tau+d}\\
\nonumber  R& \to & R|c \tau +d| \\ \nonumber  A &\to& A
e^{i\varphi}
\\
\nonumber  \mathbb{W}&\to& \begin{pmatrix} a & -b\\
                        -c & d
 \end{pmatrix}\mathbb{W}\eea

where $c\tau+d= |c\tau+d|e^{-i\varphi}$ and $\Lambda=\begin{pmatrix} a & b\\
                        c & d
 \end{pmatrix}\in Sp(2,Z)$. The hamiltonian density $\mathcal{H}$ of the compactified supermembrane is then
 invariant under (\ref{17})\cite{sl2z}. The $SL(2,\mathbb{Z})$ matrix acts from the left of the matrix $\mathbb{W}$.

\subsection{$SL(2,\mathbb{Z})$) of the Riemann surface}

When the theory is restricted by the central charge condition (the irreducible winding condition), the theory is also invariant under an extra $SL(2,\mathbb{Z})$ symmetry acting on the homology basis of the base manifold $\Sigma$, a genus one Riemann surface. This $SL(2,\mathbb{Z})$ realizes the modular transformations\footnote{In particular the supermembrane with central charges is invariant under the conformal maps homotopic to the identity.} on the upper-half plane. 
 
 
 This extra $SL(2,Z)$ symmetry is associated to the 
diffeomorphisms changing the homology basis, and consequently the
normalized harmonic one-forms, by a modular transformation on the
Teichm\"uller space of the base torus $\Sigma$. In fact, if
\bea\label{12} d\widehat{X}^{'r}(\sigma)=S^{r}_{s}
d\widehat{X}^{s}(\sigma) \eea provided \bea
\epsilon_{rs}S_{t}^{r}S_{u}^{s}=\epsilon_{tu} \eea that is $S\in
Sp(2,Z)\equiv SL(2,\mathbb{Z})$.  All
conformal transformations on $\Sigma$ are symmetries of the supermembrane with central charges \cite{sl2z}
. We
notice that under (\ref{12}) \bea \label{15}dX\to 2\pi R
(l^{'}_{s}+m^{'}_{s}\tau)d\widehat{X}^{'s}+dA^{'} \eea where
$A^{'}(\sigma^{'})=A(\sigma)$ is the transformation law of a scalar.
and the winding matrix transforms as,
\begin{equation}\label{16}
\mathbb{W}\to \mathbb{W} S^{-1}
\end{equation}

In the case of the compactified M2 ($n=0$) this  symmetry is not present in the theory since $W$ strongly depends on the central charge condition that restricts the values of the winding to those associated to the irreducible wrapping, to define the non trivial second Cohomolgy class associated to the nontrivial fibration. 

\section{T-duality in the Supermembrane Theory}
In this section we introduce the T-duality transformations for the supermembrane theory \cite{gmpr2}. This goes beyond the T-duality of superstring theory. The T-duality transformation we consider, is a nonlinear map which interchange the winding modes $\mathbb{W}$, associated to the cohomology of the base manifold with the KK charges, $Q =(p,q)$ associated to the homology of the target torus together with a transformation of the real moduli $R \to \frac{1}{R}$ and complex moduli $\tau \to \widetilde{\tau}$, both in a nontrivial way. In the following all transformed quantities under T-duality are denoted by a tilde.\newline
 In order to define the T-duality transformation we introduce the following \cite{sl2z}(47)  dimensionless variables
\bea
\mathcal{Z}:= T_{11}A\widetilde{Y}\quad \mathcal{\widetilde{Z}}:= T_{11} \widetilde{A}Y
\eea

where $T_{11}$ is the supermembrane tension, $A = (2\pi R)^2 Im\tau$ is the area of the target torus and $Y=\frac{R Im\tau}{\vert q\tau -p\vert}$. The tilde variables $\widetilde{A}, \widetilde{Y}$ are the transformed quantities under the T-duality. See (\ref{trans}) for the explicit value of $\mathcal{Z}$. The T-duality transformation we introduce is given by \cite{gmpr2}:
\bea\label{tt1}
\begin{aligned}
\textrm{The moduli}:& \quad \mathcal{Z}\widetilde{\mathcal{Z}}=1,\quad\widetilde{\tau}=\frac{\alpha \tau+\beta}{\gamma\tau +\alpha}; \\
\textrm{The charges}:& \begin{pmatrix}\widetilde{p}\\ \widetilde{q}\end{pmatrix}=\Lambda_0 \begin{pmatrix} p\\ q\end{pmatrix},
\begin{pmatrix}\widetilde{l}_1& \widetilde{l}_2\\ \widetilde{m}_1 & \widetilde{m}_2\end{pmatrix}=\Lambda_0^{-1}\begin{pmatrix}l_1^{'} & l_2^{'}\\ m_1^{'} & m_2^{'} \end{pmatrix}.
\end{aligned}
\eea
With $\Lambda_0= \begin{pmatrix}\alpha & \beta\\ \gamma & \alpha\end{pmatrix}\in SL(2,Z)$. In the above definition the T-dual supermembrane corresponds to a new supermembrane where the role of winding and KK charges interchanged, i.e. the KK modes are mapped onto the winding modes $
\begin{pmatrix}
\widetilde{p}\\ \widetilde{q}\end{pmatrix}=\begin{pmatrix}
l_1^{'}\\ m_1^{'}\end{pmatrix}$ and viceversa.
The above property together with the condition $Z\widetilde{Z}=1$ ensure that $(\textrm{T-duality})^2= \mathbb{I}$, the main property of T-duality.  The explicit transformations of the real modulus, obtained from the above T-duality transformation is
\begin{equation}\label{trans}
\widetilde{R}= \frac{\vert \gamma\tau+\alpha\vert \vert q\tau-p\vert ^{2/3}}{T_{11}^{2/3}(Im \tau)^{4/3}(2\pi)^{4/3}R},\qquad \\ \textrm{with}\quad\widetilde{\tau}=\frac{\alpha \tau+\beta}{\gamma\tau +\alpha} \quad \textrm{and}\quad \mathcal{Z}^3=\frac{T_{11} R^3 (Im\tau)^2}{\vert q\tau-p\vert}
\end{equation}
The winding modes and KK charge contribution in the mass squared formula transform in the following way:
\bea
\begin{aligned}
T_{11}n^2 A^2 &= \frac{n^2}{\widetilde{Y}^2}\mathcal{Z}^2,\qquad
\frac{m^2}{Y^2}&= T_{11}^2m^2 \widetilde{A}^2\mathcal{Z}^2.
\end{aligned}
\eea
To see how the $H_1$ (\ref{28}) transforms under T-duality it is important to realize the transformation rules for the fields,
 \bea
 \begin{aligned}
 dX^m=u d\widetilde{X}^m,\quad
 d\widetilde{X}=u e^{i\varphi}dX,\quad
 A=u e^{i\varphi}\widetilde{A}\quad
 \Psi=u^{3/2} \widetilde{\Psi},\quad
 \overline\Psi=u^{3/2}\widetilde{\overline\Psi}
 \end{aligned}
 \eea
 Where $u=\mathcal{Z}^2=\frac{R\vert \gamma\tau+\alpha\vert}{\widetilde{R}}$, $\varphi$ a phase defined in (3.22) of \cite{gmpr2} and $dX=dX^1+idX^2$ and respectively, its dual $d\widetilde{X}$ is\bea
\begin{aligned}
d\widetilde{X}=2\pi \widetilde{R}[(\widetilde{m}_1\widetilde{\tau}+\widetilde{l_1})d\widehat{X}^1+(\widetilde{m}_2\widetilde{\tau}+\widetilde{l}_2)d\widehat{X}^2]
\end{aligned}
\eea
The phase $e^{i\varphi}$ cancels with the h.c. the transformation of the Hamiltonian.
 The relation between the hamiltonians through a T-dual transformation is
 \bea
 H=\frac{1}{\widetilde{\mathcal{Z}}^{8}}\widetilde{H},\quad \widetilde{H}= \frac{1}{\mathcal{Z}^8}H.
 \eea

 We thus obtain for the mass squared formula the following identity,
 \bea
 M^2 = T_{11}^2 n^2 A^2 + \frac{m^2}{Y^2}+ T_{11}^{2/3}H =\frac{1}{\widetilde{\mathcal{Z}}^2}(\frac{n^2}{\widetilde{Y}^2}+ T_{11}^2 m^2 \widetilde{A}^2)+ \frac{T_{11}^{2/3}}{\widetilde {\mathcal{Z}}^8}\widetilde{H}.
\eea
%

\subsection{T-Duality on Symplectic Bundles}
T-duality does not only acts at local level but also globally. 
 As shown in \cite{gmpr2} one can define an equivalence class with the elements of the coinvariant group associated to the monodromy group $G$, 
$
 \{\mathcal{Q}+g\widehat{\mathcal{Q}}-\widehat{\mathcal{Q}}\},
$
such that for any $g\in G$ and $\widehat{\mathcal{Q}}\in H_1 (T^2)$, it characterizes one symplectic torus bundle. In the formulation of the supermembrane on that geometrical structure, $\mathcal{Q}$ are identified with the KK charges. The action of $G$, the monodromy group, leaves the equivalence class invariant. We now consider the duality transformation introduced previously. Under the duality transformation the equivalence class transform as

\bea
\{\mathcal{Q}+g\widehat{\mathcal{Q}}-\widehat{\mathcal{Q}}\} \to \{\Lambda_0\mathcal{Q}+(\Lambda_0 g \Lambda_0^{-1})\Lambda_0\widehat{\mathcal{Q}}-\Lambda_0\widehat{\mathcal{Q}}\},
 \eea
 hence for the dual bundle it holds,
$
 \{\Lambda_0\begin{pmatrix}l_1\\ m_1\end{pmatrix}+(\Lambda_0 g \Lambda_0^{-1})\begin{pmatrix}\widehat{l_1}\\ \widehat{m_1}\end{pmatrix}-\begin{pmatrix}\widehat{l_1}\\ \widehat{m_1}\end{pmatrix}\},$
 That is, as an element of the coinvariant group of $\Lambda_0 G \Lambda_0^{-1}$. We then conclude that the duality transformation, in addition to the transformation on the moduli $R,\tau$, also maps the geometrical structure onto an equivalent symplectic torus bundle with monodromy $\Lambda_0 G\Lambda_0^{-1}$. We notice that the transformation depends crucially on the original equivalence class of the coinvariant group. So for a nonequivalent symplectic torus bundle the dual transformations is realized with a different $SL(2,Z)$ matrix $\Lambda_0$. Now we are in position to determine the T-duality as a natural symmetry for the family of
supermembranes with central charges. We take: \bea\label{la} \widetilde{Z}=Z=1\Rightarrow
T_0=\frac{\vert q\tau-p\vert}{R^3 (Im\tau)^2}. \eea
It imposes a relation between the energy scale of the tension of the supermembrane and the moduli of
the torus fiber and that of its dual. Indeed we can think in two different ways: given the values of the moduli it fixes the allowed tension $T_0$ or on the other way around, for a fixed tension $T_0$, the radius, the Teichmuller parameter of the 2-torus, and the KK charges satisfy (\ref{la}).
\section{String Theory Limit}
We then
consider within the physical configurations of the supermembrane with central charges, the
string-like configurations \bea X^{m}=X^{m}(\tau,
q_{1}\widehat{X}^{1}+q_{2}\widehat{X}^{2}),\quad  A^{r}=A^{r}(\tau,
q_{1}\widehat{X}^{1}+q_{2}\widehat{X}^{2}) ,\eea where $q_{1},q_{2}$
are relative prime integral numbers. $X^{m}, A^{r}$ are scalar
fields on the torus $\Sigma$, a compact Riemann surface, hence they
may always be expanded on a Fourier basis in term of a double
periodic variable of that form. The restriction of $q_{1}, q_{2}$ to
be relatively prime integral numbers arises from the global
periodicity condition. On that configurations all the brackets \bea
\{X^{m},X^{n}\}=\{X^{m}, A^{r}\}=\{A^{r},A^{s}\}=0 \eea vanish.
We then obtain the final expression for the mass
contribution of the string states \cite{sl2z}:
 \bea\label{78} M_{11}^{2}\vert_{SC}= (n
T_{11}A)^2+(\frac{m}{Y})^{2}+ 8\pi^{2} R_{11}T_{11}\vert q\tau -p \vert
(N_{L}+N_{R}) \eea where $(p,q)$ are relatively prime. We notice that $(p,q)$ may be interpreted as
the wrapping of the membrane around the two cycles of the target torus.
 The corresponding change in the harmonic sector is \cite{sl2z}
\bea
dX_{h}=(qmd\widetilde{X}^{1}+pd\widetilde{X}^{2})
+\widetilde{\tau}(-Q n d\widetilde{X}^{1}+Pd\widetilde{X}^{2}),
\eea
the hamiltonian is invariant under that change. $p,q$ and $Q,P$ are now the
winding numbers of the supermembrane. Given $p,q$ there always exist
$Q$ and $P$
with the above property, although the correspondence is not unique. The $(p,q)$
type IIB strings may indeed be interpreted as different wrappings of the supermembrane with central charges.
This nice interpretation was first given in \cite{schwarz}.

The $(p,q)$ $IIB$ string compactified on a circle of radius $R_{B}$
has tension \cite{schwarz} \bea T_{(p,q)}^{2}=\frac{\vert
q\lambda_{0}-p  \vert^{2}}{Im\lambda_{0}}T^{2} \eea  where $T= T_{11}^{2/3}$ is the string tension and
$\lambda_{0}=\xi_{0}+ie^{-i\phi_{0}}$ with $\xi$ and $\phi$ 
identified with the scalar fields of the type IIB theory, $\phi$
corresponds to the dilaton fields. $\lambda_{0}$ is the asymptotic
value of $\lambda$ -the axion-dilaton of the type IIB theory-
specifying the vacuum of the theory. The perturbative spectrum of
the $(p,q)$
 IIB string is \cite{schwarz},
\bea\label{82} M_{B}^{2}=(\frac{n}{R_{B}})^{2}+(2\pi R_{B}m
T_{(p,q)})^{2}+4\pi T_{(p,q)} (N_{L}+N_{R}) .\eea If we use following
\cite{schwarz} a factor $\beta^{2}$ to identify term by term
 both mass formulas ($M_{11}=\beta M_B$), since there were obtained using different metrics, one gets
\bea\label{83} \tau= \lambda_{0},\quad  \beta^{2}=
 \frac{T_{11}A_{11}^{1/2}}{T},\quad R_{B}^{-2}= T T_{11}
A_{11}^{3/2}. \eea  
They were obtained by counting modes under some
assumptions on the supermembrane wrapping modes, as mentioned on one
of the footnotes \cite{schwarz}. Here we have derived the
expressions from a consistent definition of the supermembrane with central charges. \newline 
The identification of (\ref{78}) to the mass formula of IIA string
compactified on a circle of radius $R_{A}$ and tension $T_{A}$ may
also be performed. In order to have a consistent identification one
has to take $Re\tau=0$, $p=1$ and hence $q=0$ in (\ref{78}). The
mass formula for the perturbative spectrum of type IIA is \bea
M_{A}^{2}=(\frac{m}{R_{A}})^{2}+(2\pi R_{A}n T_{A})^{2}+4\pi T_{A}
(N_{L}+N_{R}) \eea Identification after the limit process of the winding contributions and KK ones
 using a factor $(\beta\gamma)$ to compare the $mass^{2}$ formulas,
since they are obtained using different metrics, yields
$
 R_{A}=\beta\gamma R_{11},\quad
T_{A}=\gamma^{-2}(Im\tau)^{1/2} T
$
which imply
\bea
(2\pi R_{A}R_{B})=(\frac{1}{T_{A}T_{(p,q)}})^{1/2}
\eea
We have thus obtained the $(p,q)$ IIB and IIA perturbative spectrum,
when compactified on circles $R_{B}$ and $R_{A}$ respectively, from the string states on
the supermembrane with central charges. \newline
In this limit, by restricting the worldvolume configurations of the M2 to those of the string  \cite{sl2z}, we exactly recover the mass operator of the IIB theory as was formerly conjectured by Schwarz. The gain is that from the supermembrane with central charges the pure membrane excitations are known.  If now a T-duality is performed on the supermembrane M2 mass operator restricted to string-like configurations, then an SL(2,Z) non-perturbative multiplet of IIA is obtained \cite{sl2z}.
\paragraph{String T-duality transformation limit.}
 Now we can take directly the limit of the T-duality transformations of the supermembrane to recover the standard T-duality transformations for the closed string operator. 
Before taking the limit it is convenient to consider a redefinition of $X^1$ and $X^2$. The coordinates that wrap on the $T^2$. We take 
\bea
X^1\to\frac{X^1}{T_{11}^{1/6}R_{11}^{1/2}}, \quad X^2\to T_{11}^{1/6}R_{11}^{1/2} X^2,
\eea
such that the Lie brackets $\{X^1,X^2\}^2$ will remain invariant under the redefinition. Now we consider the following limit for the torus collapsing into a circle, by imposing $ R_{11}\to 0$ . Since we want to ensure $A_{11}Y$ remains finite but $A_{11}\to 0$, necessarily $q=0$, for arbitrary $p$. Indeed this is equivalent to consider the KK charges $(p,q)=m ( 1,0)$ for $m=p$. By using the M2 T-duality transformation it can also be seen that $\tilde{A}_{11}\to 0$, so the dual also corresponds to a string. Moreover substituting in the previous definitions it can be seen  $R_A$ is finite implies $R_B$ finite.
We will define  the following for the torus degenerating into a circle $S^1$ 
\bea
R_1=\frac{R_{11}^{1/2}}{T_{11}^{1/6}}; \quad R_{2}=R_{11}^{3/2}T_{11}^{1/6} Im\tau
\eea

Since $R_{11}\to 0$ , it implies $R_1\to 0$ but $R_2$ is finite, so it corresponds to a closed curve that topologically is a circle. Now we re-express the winding condition in terms of the new variables. In terms of the new variables we get
\bea
\oint_{\mathcal{C}_s} dX^1= 2\pi R_1 l_s;\quad \oint_{\mathcal{C}_s} dX^2= 2\pi R_2 m_s,
\eea
Since $R_1\to 0$, although $l_s$ is taking finite, the first winding condition vanishes and the only residual winding condition is associated to the $S^1$ modulus is $R_2$. The former T-duality relations of the moduli in this limit become reduced to:
\bea
  Z\widetilde{Z}=1 \vert_{string},\quad \Rightarrow\quad
   T_{M2}^{4/3} R_2^3\widetilde{R}_2^3=1\to \widetilde{R}_2=\frac{\alpha^{'}}{R_2}. \eea
 where $\widetilde{R}_2= T_{11}^{1/2}\widetilde{A}^{1/2}\widetilde{Y}^{1/2}$. This defines for $T_{(p,q)}=T$ on the IIB string side precisely the duality relation of the strings.
The transformation on the charges and windings are given by (\ref{tt1}) and we finally obtain:\bea \{R;(l_1,m)\}\stackrel{T-duality}{\longrightarrow}\{\widetilde{R}=\frac{\alpha^{'}}{R}; (m,l_1)\}\eea where $m$ is the common factor between the charges.  
\section{Discussion and Conclusions}
We show that the supermembrane compactified on a torus, in both sectors have an $SL(2,Z)_{T^2}$ symmetry associated to the torus target space. The supermembrane with central charges has an extra symmetry associated to the diffeomorphisms not connected to the identity, changing the homology basis $SL(2,Z)_{\Sigma}$.
We showed the existence of a new $Z_2$ symmetry that plays the role of T-duality in M-theory interchanging the winding and KK charges but leaving the hamiltonian invariant. The supermembrane compactified on a torus realizes this duality as an exact symmetry of the theory in both sectors ($n =0$, $n\ne 0$). This is a relevant property expected for a sector of M-theory. When only string-like states are considered but it is performed this generalized T-duality of the supermembrane we obtain the mass operator IIB and the corresponding dual realizes the type IIA with all nonperturbative $SL(2,Z)$ multiplet. If we take the limit of the T-duality transformation of the supermembrane into the standard one, then only the standard $(p,q)= (1,0)$ mass type IIA operator is allowed meanwhile IIB mass operator is unchanged.
The Supermembrane Theory compactified on a torus. Supermembrane theory realize S, T so U-duality as exact symmetries of the theory. There are two well defined sectors: without and with central charges. While the first one has as U-duality group in 9D as symmetry the group $SL(2,Z)_{T^2}\times Z_2$, the second one exhibits a larger duality group,  $SL(2,Z)_{\Sigma}\times SL(2,Z)_{T^2}\times Z_2$ symmetry.
 We have been able to prove that U-dualities are symmetries of a toroidally compactified sector of M-theory, the supermembrane on a torus, as was already conjectured in \cite{dewit}.
 
\section{Acknowledgements}I specially thanks A. Restuccia for collaboration, helpful comments and kind support and to my collaborators J.M. Pena and I. Martin together with A. Restuccia, with whom these results were obtained. Part of the work of MPGM was funded by the Spanish
Ministerio de Ciencia e Innovaci\'on (FPA2006-09199) and the
Consolider-Ingenio 2010 Programme CPAN (CSD2007-00042). MPGM also thanks to Proyecto FONDECYT 1121103.

\end{document}